\documentclass{aastex63}
	
\usepackage{graphicx}
\usepackage{amsmath}
\usepackage{longtable}
\usepackage{color}
\usepackage{enumerate}

\received{XXX}
\revised{YYY}
\accepted{ZZZ}

\submitjournal{ApJ}

\shorttitle{Precession model for FRB 180916}
\shortauthors{Chen et al.}

\begin{document}

\title{Reconciling the 16.35-day period of FRB 20180916B with jet precession}

\correspondingauthor{Wei-Min Gu; Mouyuan Sun}
\email{guwm@xmu.edu.cn; msun88@xmu.edu.cn}

\author{Hao-Yan Chen}
\affiliation{Department of Astronomy, Xiamen University, Xiamen,
Fujian 361005, P. R. China}

\author{Wei-Min Gu}
\affiliation{Department of Astronomy, Xiamen University, Xiamen,
Fujian 361005, P. R. China}

\author{Mouyuan Sun}
\affiliation{Department of Astronomy, Xiamen University, Xiamen,
Fujian 361005, P. R. China}

\author{Tong Liu}
\affiliation{Department of Astronomy, Xiamen University, Xiamen,
Fujian 361005, P. R. China}

\author{Tuan Yi}
\affiliation{Department of Astronomy, Xiamen University, Xiamen,
Fujian 361005, P. R. China}

\begin{abstract}

A repeating fast radio burst (FRB), FRB 20180916B (hereafter FRB 180916), was reported to have a 16.35-day period. This period might be related to a precession period. 
In this paper, we investigate two precession models
to explain the periodic activity of FRB 180916.
In both models, the radio emission of FRB 180916 is produced by a precessing jet. 
For the first disk-driven jet precession model,
an extremely low viscous parameter (i.e., the dimensionless viscosity parameter $\alpha \lesssim 10^{-8}$) is required to explain the precession of FRB 180916, which implies its implausibility. For the second tidal force-driven jet precession model,
we consider a compact binary consists of a neutron star/black hole and a white dwarf;
the white dwarf fills its Roche lobe and mass transfer occurs.
Due to the misalignment between the disk and orbital plane, the tidal force of the white dwarf can drive jet precession. We show that the relevant precession periods are
several days to hundreds of days, depending on the specific accretion rates and component masses. The duration of FRB 180916 generation in the binary with extremely high accretion rate will be several thousand years.

\end{abstract}

\keywords{Compact binary stars (283)---Jets (870)---Stellar accretion disks (1579)---White dwarf stars (1799)---Radio transient sources (2008)}

\section{Introduction} \label{sec:intro}

Fast radio bursts (FRBs) are mysterious millisecond-duration radio bursts and  have been detected in frequencies from 300 MHz \citep{Chawla et al.(2020)}
to 8 GHz \citep{Gajjar et al.(2018)}.
Recently, \citet{Pastor-Marazuela et al.(2020)} and \citet{Pleunis et al.(2021)} observed FRB 20180916B (hereafter FRB 180916) down to 120 MHz.
Since the dispersion measure (DM) is more than the contribution of the
Galaxy, FRBs are thought to be extragalactic sources
\citep{Petroff et al.(2019),2019ARA&A..57..417C, Platts et al.(2019)}.
Some host galaxies of FRBs have been identified at redshifts
in the range $z \approx 0.03-0.66$ \footnote{\url{http://frbhosts.org}},
which strongly supports the extragalactic nature of FRBs \citep[e.g.,][]{Tendulkar et al.(2017)}.
On April 28, 2020, however, a luminous millisecond radio burst from SGR 1935+2154 (FRB 200428) was observed in the Galaxy \citep{Bochenek et al.(2020), CHIME/FRB Collaboration et al.(2020b), Lin et al.(2020), Li et al.(2021)}. 
Unlike most FRBs, FRB 200428 has two $\sim$ 5-ms long sub-bursts 
separated by $\sim$ 30 ms \citep{Ridnaia et al.(2021)}.
The number of outbreaks can be used to classify FRBs into two categories, i.e., repeating FRBs and one-off ones.
The repeating and non-repeating FRBs differ in several aspects. For example,
the repeating FRBs have longer pulse widths and weaker radio luminosities than the non-repeating ones.  Meanwhile, the ``down-drifting'' of frequency and multiple 
sub-bursts, which seem to be common in the repeating FRBs \citep
{CHIME/FRB Collaboration et al.(2019),Fonseca et al.(2020),Cui et al.(2021)}, 
are absent in non-repeating ones.

\citet{CHIME/FRB Collaboration et al.(2020a)} reported the first periodic
activity of FRBs, i.e., the source FRB 180916 exhibits an activity period of
16.35 days and has a $\sim$ 5-day burst window. After that, FRB 121102 was also found to have a $\sim$ 157-day activity period with a duty cycle of 56 $\%$
\citep{Rajwade et al.(2020),Cruces et al.(2021)}. The discovery of periodic 
activities provides important clues for revealing the physical mechanisms of 
repeating FRBs. 
It is proposed that the activity period of FRB 180916 might be related to either the orbital or the precession period \citep{CHIME/FRB Collaboration et al.(2020a)}. 
For the orbital period scenario, models involve a neutron star (NS) and white dwarf (WD) binary \citep{Gu et al.(2020)}, a pulsar passes through the
asteroid belt \citep{2020ApJ...895L...1D}, and a mild pulsar in tight O/B star
binary \citep{Lyutikov et al.(2020)} are suggested. 
For the precession scenario, models often invoke the
precession of the NS in a black hole (BH)-NS binary system
\citep{2020ApJ...893L..31Y}, or the free precession of a magnetar
\citep{Levin et al.(2020),Tong et al.(2020), 2020ApJ...892L..15Z}. Note that
due to the fast damping caused by the pinning of superfluid vortices inside
the star, the long-life free precession of an isolated NS is
considered impossible \citep{CHIME/FRB Collaboration et al.(2020a)}.
There are many other difficulties in the NS model including the energy budget
and the strength of the magnetic field \citep{Katz(2020)}. 
\citet{Katz(2017)} suggested that an intermediate-mass BH with chaotic
accretion (i.e., the accreting gas has random angular momentum) can produce a 
wandering jet and the repeating FRBs without
periodicity are produced by this jet.
Using the disk-driven jet precession model originally proposed
by \citet{Sarazin et al.(1980)}, \citet{Chen(2020)} proposed a young pulsar
with debris disk precession to explain the periodic activity of FRB 180916.
Then, \citet{Katz(2021a)} proposed some tests to distinguish different periodic FRB models of the 16.35-day periodicity of FRB 180916.

Precession might be a common phenomenon in the Universe. Some active galactic nuclei
(AGNs) have long-timescale precession periods, for instance, 3C 196 might have a $10^{6}$-year precession period \citep{2004MNRAS.349.1218C}. 
Moreover, some X-ray binaries also precess and have super-orbital periods
(precession periods). For instance, the 162.5-day
\citep[e.g., ][]{Katz(1980),Sarazin et al.(1980),Katz et al.(1982)}
and 35-day \citep{Katz(1973)} super-orbital periods of SS 433 and Her X-1
are attributed to the precession period. 
In many binary systems, SS 433 is the most well-studied 
microquasar in our Galaxy. This X-ray binary contains a compact primary star and
an evolved A-type supergiant with a 13.1-day orbital period. A pair of 
well-collimated jets with velocity $\pm$ 0.26c repeat the 162.5-day precession
motion. The accretion rate of SS 433 is
super-Eddington as $\sim 10^{-4} \ M_{\sun} \ \rm{yr^{-1}}$ and the secondary
star fills its Roche lobe \citep{Fabrika(2004)}. Some models have been 
proposed to interpret the precession of SS 433, such as the 
accretion disk-driven BH and jet precession model
\citep{Sarazin et al.(1980)}, irradiation-driven precession model 
\citep{Begelman et al.(2006)}, and the tidal force of the companion star-driven
precession model \citep{Leibowitz(1984), Larwood(1998),Li et al.(2020)}.
However, for the irradiation-driven precession model, owing to the absence of 
external torques, the precession of the disk caused by radiation should be 
prograde, which would contradict the observation of SS 433
\citep{Maloney et al.(1998)}.
The disk-driven precession model and the tidal force-driven precession model
both can cause retrograde precession \citep{Caproni et al.(2006)}. 

Ultra-compact X-ray binaries (UCXBs) \citep{2010NewAR..54...87N} can also be observed the precession phenomenon in the Galaxy. For example, 47 Tuc X9, a BH-WD binary, has a 6.8-day precession period \citep{Bahramian et al.(2017)}, and X1916-053, an NS-WD binary, has a 3.8-day precession period. The accretion 
rates of such systems are lower than the Eddington accretion rate. However, 
in the state of the unstable and extremely super-Eddington accretion, a compact binary system composed of a BH and a WD ($M_{\rm{WD}}>0.6 \ M_{\sun}$) can produce long GRBs \citep{Dong et al.(2018)}. 
As the accretion rate declines, the BH-WD binaries might evolve to UCXBs. 

In this paper, we first show that the jet precession model of
\citet{Sarazin et al.(1980)},
i.e., the central engine is a stellar-mass black hole or
an intermediate-mass black hole, requires an extremely low dimensionless viscous parameter ($\alpha \la 10^{-8}$); hence we argue that this model
is implausible. Then, we propose a compact binary system with a circular orbit,
where the primary star is a BH or an NS and the secondary is a WD
($M_{\rm{WD}}<0.6 \ M_{\sun}$)
\footnote{Most recently, NS-WD binaries as repeating FRB central engines were also suggested by \citet{Katz(2021b)}}.
The WD fills its Roche lobe and the mass transfer occurs through the Roche-lobe overflow. 
The system has a stable and super-Eddington accretion rate. Due to the high accretion rate, a narrowly collimated jet can be formed. The disk and jet will precess together with a
16.35-day precession period. Different from the precession of the NS caused
by the coupling of the orbit and spin of the BH-NS binary \citep{2020ApJ...893L..31Y}, we consider that the precession of the jet is driven by the tidal
force of WD. FRBs can be observed when the jet sweeps across
the direction of the observers. 

The remainder of this paper is organized as follows. In Section \ref{sec:model},
we test the precession models for the periodic activity of FRB 180916,
i.e., the accretion disk-driven jet precession model and the tidal
force-driven jet precession model. In Section \ref{sec:system}, we discuss the properties
of the binary system. 
Conclusions and discussion are presented in Section \ref{sec:con}.

\section{The Model} \label{sec:model}

\subsection{Accretion Disk-Driven Jet Precession} \label{subsec: jet}

We take SS 433 as an example to introduce the accretion disk-driven jet
precession model. SS 433 is an X-ray
binary, containing an $8 \ M_{\sun}$ spinning BH and a $24 \ M_{\sun}$ 
nondegenerate star \citep{2020ApJ...896...34H}. 
In this model, it is assumed that the outer regions of the accretion disk are
tilted to the orbital plane. The spin axis of the BH and the binary
axis are misaligned. The inner regions of the accretion disk and BH are gravitationally coupled and can precess together due to the
Lense-Thirring effect \citep{Sarazin et al.(1980)}. If the precession period is
shorter than the viscous timescale in the disk, the precession process
is approximate to the conservation of angular momentum. Following the spirit of 
\citet{Sarazin et al.(1980)}, we assume a ring in the disk with the width
$dr$ has an angular momentum per logarithmic interval of radius
$J(r) = dJ / d (\ln{r}) = 2\pi r^{3} \Sigma v_{\phi}$, where
$\Sigma$ is the surface density of the disk and $v_{\phi}$ is the rotational
velocity of the disk. Meanwhile, the angular momentum of the BH is $J_{\ast}= a_{\ast}G M^{2}_{\ast}/c $, where $M_{\ast}$ is the BH mass, $c$ is the speed of light, $G$  is the gravitational constant, and 
$a_{\ast}$  is the dimensionless specific angular momentum of BH, 
$0 \leqslant a_{\ast}< 1$.
Then, there exists a critical radius $r_{\rm{prec}}$ (the precession radius) 
such that $J(r_{\rm{prec}}) = J_{\ast}$. The outer portions of the disk 
($r>r_{\rm{prec}}$) with more angular momentum than the central BH can cause
the precession of the BH and the central parts of the disk ($r \ll r_{\rm{prec}}$).
If the disk is adequately massive ($10^{-5}-10^{-2} \ M_{\sun}$ \citep{Sarazin et al.(1980)}), the outer portions of the disk with sufficient
angular momentum will maintain their orientation. 
Part of the accreted inner region materials are supposed to be ejected out along the BH spin axis, forming a precessing jet. 

The precession frequency is $\Omega_{\rm{prec}}=2GJ(r)/c^{2}r^{3}$
\citep{Sarazin et al.(1980)}. Since $J(r)/r^{3}$ decreases with increasing $r$
in the outer disk, the fastest precession rate is produced at the radius
$r=r_{\rm{prec}}$ with $J(r_{\rm{prec}})=J_{\ast}$:

\begin{equation}
\Omega_{\rm{prec}}=\frac{2GJ_{\ast}}{c^{2}r^{3}_{\rm{prec}}}\ .
\end{equation}

\begin{figure}[htbp]
\centering
\includegraphics[height=10cm,width=10cm]{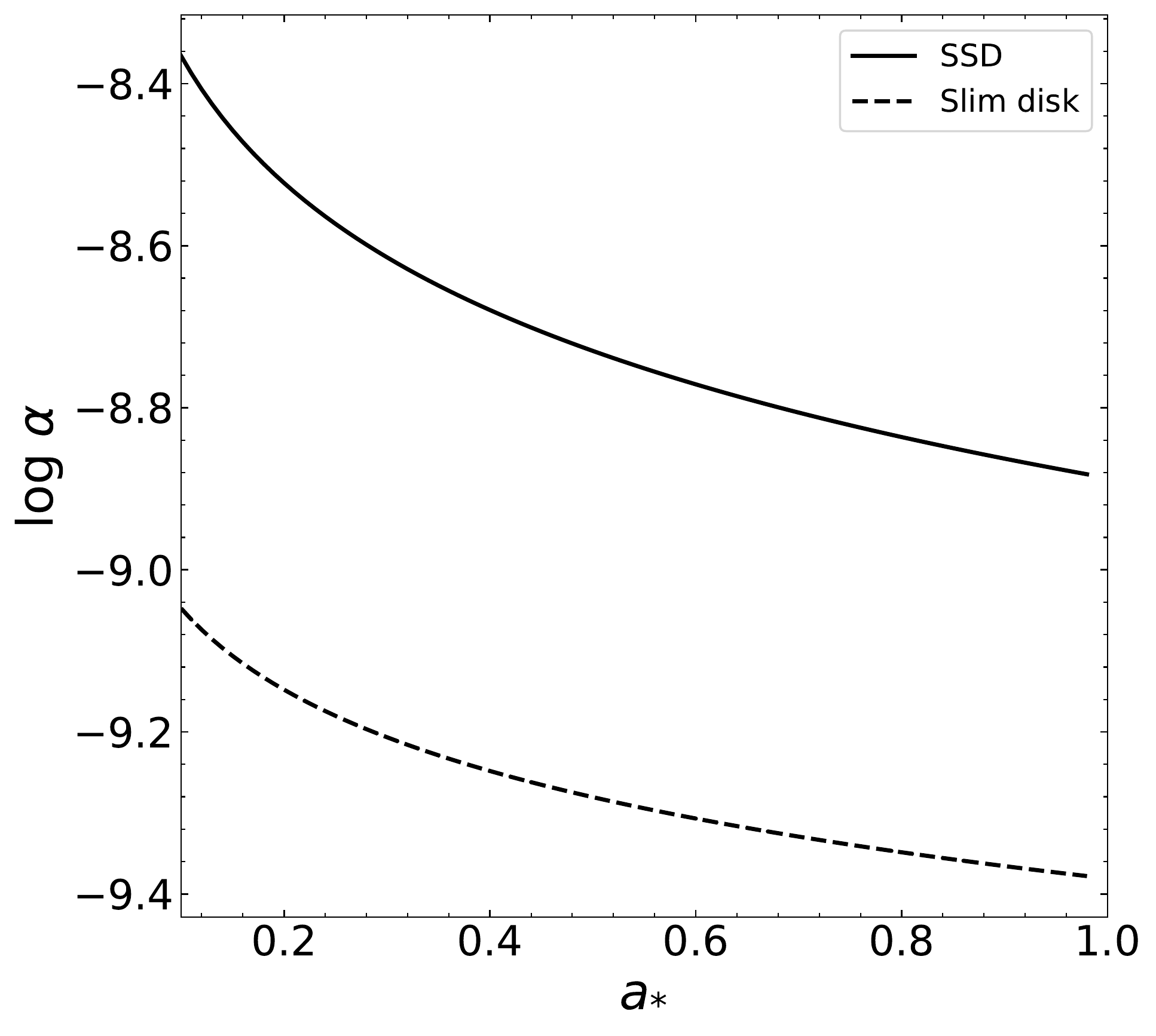}
\caption{The relationship between the value of viscous parameter $\alpha$ and
the dimensionless specific angular momentum $a_{\ast}$ of BH in different disk
models for SS 433 in a 162.5-day precession period. The solid line represents 
the standard disk model (SSD), and the dashed line represents the slim disk. 
Here the mass of BH for SS 433 is
$M_{\ast}=8\ M_{\sun}$ and the accretion rate of SS 433 in the outer disk 
is $\dot{M} \sim 10^{-4}\ M_{\sun} \ \rm{yr}^{-1}$.
The value of the viscous parameter $\alpha$ decreases with
increasing $a_{*}$ in any disk models.
\label{fig1} }
\end{figure}

By using the standard disk model (SSD) of \citet{1973A&A....24..337S},
the precession radius $r_{\rm{prec}}$ and the precession period $P_{\rm{prec}}$
can be derived as
\citep[also see Equations (2) and (5) in][]{Sarazin et al.(1980)}

\begin{equation}
r_{\rm{prec}}=2.3\times 10^{14} a^{4/7}_{\ast} \alpha^{16/35} \left(\frac{M_{\ast}}{M_{\sun}}\right)^{5/7}\left(\frac{\dot{M}_{\ast}}{10^{-8}M_{\sun} \ \rm{yr}^{-1}}\right)^{-2/5}
\ \rm{cm} \ ,
\end{equation}

\begin{equation}
P_{\rm{prec}}=7.4\times 10^{17} a^{5/7}_{\ast} \alpha^{48/35} \left(\frac{M_{\ast}}{M_{\sun}}\right)^{1/7}\left(\frac{\dot{M}_{\ast}}{10^{-8}M_{\sun} \ \rm{yr}^{-1}}\right)^{-6/5}
\ \rm{day} \ ,
\label{con: e5}
\end{equation}
where $\alpha$ is the dimensionless viscous parameter. Note that, $10^{-8} \ M_{\sun} \ \rm{yr^{-1}}$ is the Eddington accretion rate of a $1 \ M_{\sun}$ compact object. Recalling that $M_* = 8 \ M_\odot$ and $\dot{M_{\ast}} \sim 10^{-4} \ M_{\sun} \ \rm{yr^{-1}}$, according to Equation (\ref{con: e5}), the required $\alpha$ to reconcile the observed precession period is $\sim 10^{-8}$ \citep[see the solid line in Figure \ref{fig1}; also see][]{Sarazin et al.(1980)}
as long as $a_{*}>0.1$; this value is seven orders of magnitude smaller than the value inferred from the observations \citep[e.g., the AGN optical variability; see][]{Sun et al.(2020)}.
The reasons for an extremely low viscous parameter are as follows.
The outer regions of the disk should have enough angular momentum to force the BH and
the inner disk to precess. To obtain enough angular momentum, a high surface density $\Sigma$ of the disk is needed in the model. 
For fixed accretion rate, according to the mass conservation equation $\dot{M}_{\ast}=-2\pi r \Sigma v_{r}$, the radial velocity $v_{r}$ should be extremely small in the outer portions of the disk. The required slow radial velocity, which scales as $v_r\propto \alpha^{4/5}$ \citep{1973A&A....24..337S}, indicates that $\alpha$ should also be unusually small. 

Given the fact that the accretion rate in SS 433 is extremely super-Eddington, the application  of SSD to this system might not be self-consistent. 
The slim disk model \citep{Abramowicz et al.(1988)} is likely a better choice. 
In the slim disk model, the flow with $v_{\phi}$ is sub-Keplerian 
$v_{\phi} \sim V_{\rm{K}}/ \sqrt{5}$ \citep{1999ApJ...516..420W}. The radial 
velocity of the disk material $v_{r}$ is roughly equal to
$-\alpha V_{\rm{K}} / \sqrt{5}$. Thus, the precession radius and the precession
period can be expressed as

\begin{equation}
r_{\rm{prec}}=4.7\times 10^{15} a^{1/2}_{\ast} \alpha^{1/2} \left(\frac{M_{\ast}}{M_{\sun}}\right) \left(\frac{\dot{M}_{\ast}}{10^{-8}M_{\sun} \ \rm{yr}^{-1}}\right)^{-1/2}
\ \rm{cm}  ,
\end{equation}

\begin{equation}
P_{\rm{prec}}=1.7\times 10^{21} a^{1/2}_{\ast} \alpha^{3/2} \left(\frac{M_{\ast}}
{M_{\sun}}\right)\left(\frac{\dot{M}_{\ast}}{10^{-8}M_{\sun} \ \rm{yr}^{-1}}\right)^{-3/2}
\ \rm{day} \label{con: e7} .
\end{equation}

From Equation (\ref{con: e7}), the viscous parameter $\alpha$ at precession
radius $r_{\rm prec}$ for the 162.5-day precession period of SS 433 is $\sim10^{-9}$ 
(see the dashed line in Figure \ref{fig1}). According to Equation (6) of \citet{Sarazin et al.(1980)}, it is quite straightforward to see that the precession period $P_{\rm{prec}}$ depends upon the ratio of the radial to angular velocities (for fixed accretion rate). In the slim disk model, the ratio of $v_{\rm{r}}$ to $v_{\rm{\phi}}$ is larger than that in SSD. In order to produce the same precession period, for fixed accretion rate $\dot{M}_{\ast}$, the required $\alpha$ for the slim disk model should be smaller than that for SSD. 

Although the disk-driven jet precession is unlikely to account for the observed periods for SS 433, this model is able to interpret the long timescale precession of AGNs
\citep[e.g.,][]{Lu(1990),2005ApJ...635L..17L} and the burst timescale of GRBs \citep[e.g.,][]{Liu et al.(2010),Sun et al.(2012),Hou et al.(2014)}.

\citet{Katz(2017)} proposed that the repeating FRBs can be produced by
a wandering jet based on the chaotic accretion (i.e., the accreting gas has random angular momentum) in an intermediate-mass ($\thicksim 10^{2}-10^{6} \ M_{\sun}$) BH accretion system.
A repeating FRB is observed when the jet repeatedly sweeps across our line of sight.
In addition, for the
disk-driven jet precession in an intermediate-mass BH system,
an extremely low viscous parameter ($\alpha \approx 10^{-12}$ for $\dot{M} \approx 10^{-7} \ M_{\sun} \ \rm{yr}^{-1}$; $\alpha \approx 10^{-8}$ for $\dot{M} \approx 10^{-1} \ M_{\sun} \ \rm{yr}^{-1}$) is necessary to account for the 16.35-day period of FRB 180916 regardless of the BH spin, as implied by Equations (\ref{con: e5}) and (\ref{con: e7}).
Therefore, whether the central BH is a stellar mass or an intermediate mass, the disk-driven jet precession model is unlikely to account for the periodic activity of FRB 180916.

\subsection{Tidal Force-Driven Jet Precession} \label{subsec: tidal}

Another mechanism involves the jet precession of a binary system due to the tidal force of the companion star. \citet{Katz(1973)} first proposed this model to explain the precession
period of Her X-1. Then, \citet{Katz(1980)} and \citet{Leibowitz(1984)} used this model to explain the precession period of SS 433. The model can explain the
precession of binary system of T-Tauri stars and some X-ray binary systems, such as Cyg X-1 and LMC X-1 \citep[e.g.,][]{1995MNRAS.274..987P,Larwood(1997),Larwood(1998),Caproni et al.(2006)}.

If the observed period is indeed caused by the tidal force-driven jet precess, we might immediately infer some appropriate characteristics of FRB 180916. 
\citet{Larwood(1998)} and \citet{Caproni et al.(2006)} showed 
that the ratio of the precession period $P_{\rm{prec}}$ and orbital period $P_{\rm{orb}}$ decreases with increasing mass ratio of two stars in the tidal force-driven jet precession model. For SS 433, the mass ratio of two stars is $q=M_{\rm{2}}/M_{\ast}=3$ \citep{2020ApJ...896...34H}
and the period ratio is $P_{\rm{prec}}/P_{\rm{orb}} > 10$, where $M_{\rm{2}}$ is the mass of the secondary star. 
Considering the period ratio of SS 433 and the observed 16.35-day precession period of FRB 180916, we can infer that the orbital period of the
binary system for FRB 180916 is less than one day since the relevant $q$ is unlikely to be 
larger than 3. 
That is, FRB 180916 may come from such a system with super-Eddington accretion rate,
relativistic jets, and a short orbital period ($<$1 day). 
Under the condition of stable mass transfer of the binaries, the orbital periods of most double neutron star systems (DNSs) are more than one day \citep{Tauris et al.(2017)} and the accretion rates of double white dwarfs (DWDs) and BH-NS binaries are much less than the Eddington accretion rates \citep{Marsh et al.(2004)}.
Hence, we rule out the DNSs, DWDs, and BH-NS models. 
According to the above speculations, the system of FRB 180916 can be either an NS-WD binary \citep{Gu et al.(2016), Gu et al.(2020)} or a BH-WD binary \citep{Dong et al.(2018)}.
Note that the mass ratio of a BH/NS-WD binary $q\ll 3$.

\begin{figure}
\centering
\includegraphics[width=1.0\linewidth]{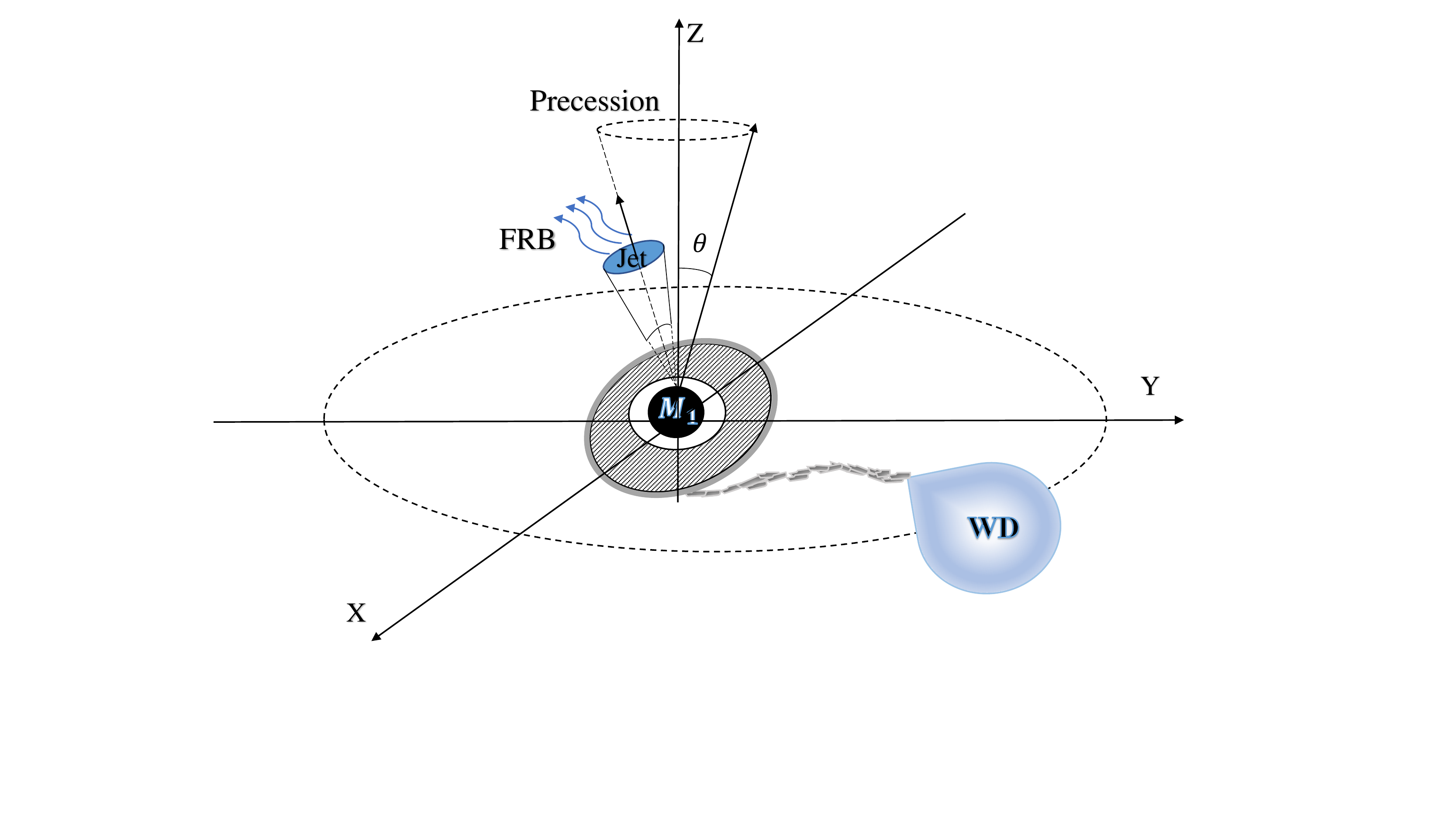}
\caption{The schematic diagram of the tidal force-driven jet precession model.
In the binary system with a circular orbit, the primary star ($M_{1}$)
may be an NS or a BH and the secondary star is a WD. The WD fills its
Roche lobe and mass transfer occurs from the WD to the primary star
through the inner Lagrangian point $L_{1}$. Owing to the tidal force
of the WD, the jet can precess along with the primary star.
Periodic FRBs can be observed when the jet periodically sweeps across
the observers. (The sizes of compact objects are not to scale for
the purpose of demonstration)
\label{fig:schematic}}
\end{figure}

The schematic diagram of our model is shown in Figure \ref{fig:schematic}.
In this model, we consider the binary system in a circular orbit. We denote
the mass of the central compact star is $M_{1}$ and the mass of the companion star
(WD) is $M_{2}$ ($M_{\rm{WD}}$).  The central object is surrounded by a
non-self-gravitating and axisymmetric disk. The outer regions
of the disk are tilted with respect to the orbital plane whereas the inner
regions of the disk are coplanar with orbit. The spin axis of the central object
is misaligned with the binary axis. In such a system, the mass
transfer occurs when the WD fills its Roche lobe. These accreted materials form
a ring at the edge of the outer portions of the central star's disk. Then, the
plane of the ring will precess due to the tidal force of the WD
\citep{Katz(1973)}. The tidal torque applied to the outer edge of the
accretion disk will transfer inward by bending wave transport or by viscous
stresses, or by a combination of both effects \citep{Larwood(1998)}. 
There is a gradient of the tilt angle in the disk with different radii, which
leads to the precession of the whole disk. Owing to the Bardeen-Petterson effect \citep{1975ApJ...195L..65B}, the central object along with the central
part of the disk precesses together. Part of the accreted materials from
the WD is supposed to be ejected out along the BH spin axis, forming
a precessing jet. 

The precession frequency $\Omega_{\rm{prec}}$ of the disk induced by the tidal force is taken the form \citep{1995MNRAS.274..987P,Larwood(1997)}

\begin{equation}
\Omega_{\rm{prec}}=- \frac{3}{4}\frac{G M_{\rm{WD}}}{a^{3}} \frac{\int \Sigma r^{3} 
dr}{\int \Omega \Sigma r^{3} dr } \cos{\theta} \label{con:period ratio},
\end{equation}
where $\Sigma$ is axisymmetric surface density, $\Omega$ is Keplerian
angular velocity and $a$ is the distance between two stars. The parameter $\theta$
indicates the orbital inclination with respect to the plane of the accretion
disk. In the circular orbit, the jet forms a precession cone with
half opening angle equal to the angle of orbital inclination $\theta$ 
\citep{2004MNRAS.349.1218C}.
The minus sign indicates that the precession is retrograde. That is,
the precession is opposite to the rotation direction of the accretion disk.

Following \citet{Larwood(1997)}, in order to avoid the detailed investigation on the energy balance, a simple polytropic relation between
gas pressure $P$ and density $\rho$ is adopted, i.e., $P = K \rho^{1+1/n}$,
where $n$ is the polytropic index.
Such a polytropic relation was widely used in literatures on accretion systems, such as the Bondi accretion and the vertical structure of accretion disks \citep[e.g., Chapter 2.2.1 and 7.2.1 of][]{Kato et al.(2008)}.
By integrating Equation (\ref{con:period ratio}) from the inner radius
$R_{\rm{in}}$ of the disk to the outer radius $R_{\rm{out}}$ of the disk , 
we can obtain \citep{Larwood(1998)} 

\begin{equation}
\frac{\Omega_{\rm{prec}}}{\Omega_{\rm{out}}}=- \frac{3}{4} \frac{7-2n}{5-n}q \left(\frac{R_{\rm{out}}}{a}\right)^{3} \cos{\theta},
\end{equation}
where $q$ is mass ratio defined as $q\equiv M_{\rm{WD}}/M_{1}$ ,
and $\Omega_{\rm{out}}$ is the Keplerian angular velocity at the outer radius
$R_{\rm{out}}$. For $n=3/2$, the precession period is written as
\citep{Larwood(1998)}

\begin{equation}
\frac{P_{\rm{orb}}}{P_{\rm{prec}}}= \frac{3}{7} q \left(\frac{1}{1+q}\right)^{1/2}\left(\frac{R_{\rm{out}}}{a}\right)^{3/2} \cos{\theta}, \label{con: Porb/Pprec}
\end{equation}
where $P_{\rm{orb}}$ is the orbital period. The size of the accretion disk is a
fraction of the primary star's Roche-lobe radius $R_{\rm{L1}}$, which is defined
as $R_{\rm{out}}=\beta R_{\rm{L1}}$. The Roche-lobe radius of the primary
star can be expressed as \citep{Eggleton(1983)}

\begin{equation}
\frac{R_{\rm{L1}}}{a}=\frac{0.49 q^{-2/3}}{0.6 q^{-2/3}+\ln\left({1+q^{-1/3}}\right)}.
\label{con:RL1/a}
\end{equation}
Thus, Equation (\ref{con: Porb/Pprec}) can be rewritten as

\begin{equation}
\frac{P_{\rm{orb}}}{P_{\rm{prec}}}= \frac{3}{7} \beta^{3/2}q
\frac{ \mathcal{R}^{3/2}\cos{\theta}}{\left(1+q\right)^{1/2}}, \label{con: Porb/Pprec 2}
\end{equation}
where $\mathcal{R}=R_{\rm{L1}}/a$. From Equation (\ref{con: Porb/Pprec 2}),
the precession period $P_{\rm{prec}}$ can be derived if the values of $\beta$, the mass ratio $q$, the orbital inclination
$\theta$, and the orbital period $P_{\rm{orb}}$ are given. \citet{Paczynski(1977)}
calculated the value of $\beta$ for mass ratio $0.03< q <30$. For mass ratio
$0.03<q<2/3$, $\beta$ is a mean value of 0.86. For mass ratio $2/3<q<30$,
$\beta$ is a function of mass ratio $q$ 

\begin{equation}
\beta (q)= \frac{1.4}{1+[\ln ({1.8q})]^{0.24}}.
\end{equation}
In our model, the binary system should keep the state of stable accretion 
and gravitational balance, so the mass ratio should $q< 2/3$ \citep{King et al.(2007a)}. 
Consequently, $\beta$ can take a constant value of 0.86. 
Here we take an average half opening angle of the jet about $\theta=20{\degr}$ \citep{Caproni et al.(2006),Larwood(1998)}.

The WD fills its Roche lobe, i.e., the radius of the Roche lobe
is equal to the radius of WD, $R_{\rm{L2}}=R_{\rm{WD}}$.
$R_{\rm{WD}}$ can be expressed as \citep{1988ApJ...332..193V}

\begin{equation}
R_{\rm{WD}}=0.0114 \ R_{\sun} \left[\left(\frac{M_{\rm{WD}}}{M_{\rm{Ch}}}\right)^{-2/3}-
\left(\frac{M_{\rm{WD}}}{M_{\rm{Ch}}}\right)^{2/3}\right]^{1/2} \times
\left[1+3.5\left(\frac{M_{\rm{WD}}}{M_{\rm{p}}}\right)^{-2/3}+\left(\frac{M_{\rm{WD}}}{M_{\rm{p}}}\right)^{-1}\right]^{-2/3} \ ,\label{con:Rwd}
\end{equation}
where $M_{\rm{Ch}}$ is the Chandrasekhar mass limit 
$M_{\rm{Ch}}=1.44 \ M_{\sun}$ and  $M_{\rm{p}}=0.00057 \ M_{\sun}$. The radius 
of the Roche lobe for WD $R_{\rm{L2}}$ can be expressed as 
\citep{Eggleton(1983)}

\begin{equation}
\frac{R_{\rm{L2}}}{a}=\frac{0.49 q^{2/3}}{0.6 q^{2/3}+\ln({1+q^{1/3}})}. 
\label{con:RL2/a}
\end{equation}
The Kepler's third law for this system is 

\begin{equation}
\frac{G(M_{1}+M_{\rm{WD}})}{a^{3}}= \frac{4\pi^{2}}{P^{2}_{\rm{orb}}}. 
\label{con:third law}
\end{equation}
Once $M_{1}$ and $M_{\rm{WD}}$ are given, we can obtain $P_{\rm{orb}}$ 
and $a$ by Equations (\ref{con:Rwd})-(\ref{con:third law}). Then, we can infer the precession period $P_{\rm{pre}}$ by Equation(\ref{con: Porb/Pprec 2}).

As a second step, we estimate the mass transfer rate of the binary system.
SS 433 has a pair of precessing jets with a super-Eddington accretion rate. 
We consider that our model should also have high accretion rates in order to 
produce the collimated jets.
\citet{Dong et al.(2018)} proposed a BH-WD binary system with an
extremely super-Eddington accretion rate which has violent accretion
to explain long GRBs without detection of supernova association. 
On the contrary, under the case of stable accretion, the mass of BH/NS-WD binary transfers through Roche-lobe overflow. The mass transfer
rate from the WD to its companion is \citep{Dong et al.(2018)}

\begin{equation}
\dot {M} = - \dot{M}_{\rm{WD}} =
- \frac{64 G^{3} M_{1}M^{2}_{\rm{WD}}M}{5 c^{5}a^{4}
\left(2q-\frac{5}{3}+\frac{\delta}{3}\right)},
\label{con: accretion}
\end{equation}

\begin{equation}
\delta=\frac{\left(\frac{M_{\rm{WD}}}{M_{\rm{Ch}}}\right)^{-2/3}+\left(\frac{M_{\rm{WD}}}
{M_{\rm{Ch}}}\right)^{2/3}}{\left(\frac{M_{\rm{WD}}}{M_{\rm{Ch}}}\right)^{-2/3}-
\left(\frac{M_{\rm{WD}}}{M_{\rm{Ch}}}\right)^{2/3}}-\frac{2\frac{M_{\rm{p}}}{M_{\rm{WD}}}
\left[\frac{7}{3}\left(\frac{M_{\rm{p}}}{M_{\rm{WD}}}\right)^{-1/3}+1\right]}{1+3.5\left(\frac{M_{\rm{p}}}
{M_{\rm{WD}}}\right)^{2/3}+\left(\frac{M_{\rm{p}}}{M_{\rm{WD}}}\right)}, 
\label{con: theta}
\end{equation}
where $M=M_{1}+M_{\rm{WD}}$ is the total mass of the primary and secondary star.
Then, a dimensionless parameter, specific accretion rate, is defined as
$\dot{m} = \dot{M}/\dot{M}_{\rm{Edd}}$, where
$\dot{M}_{\rm{Edd}}=10L_{\rm{Edd}}/ c^{2}$ is the Eddington accretion rate
for the primary star. Thus, 

\begin{equation}
\dot{m}=\frac{4G^{2}M^{2}_{\rm{WD}}M \kappa_{\rm{es}}}{25\pi c^{4}a^{4}
\left(\frac{5}{6}-q-\frac{\delta}{6}\right)}, \label{con: dot{m}}
\end{equation}
where the opacity $\kappa_{\rm{es}}=0.34 \ \rm{cm}^{2}\ \rm{g}^{-1}$. 

We now need to evaluate the appropriate
accretion rate for FRB 180916 in BH/NS-WD binary system. 
The isotropic radio luminosities of FRBs are estimated to be in a range 
$L_{\rm{radio}} \ \sim 10^{38}-10^{42} \ \rm{erg} \ \rm{s}^{-1}$ 
\citep{2019ARA&A..57..417C}. However, due to our poor knowledge of radiative efficiency $\eta$ and the bolometric luminosity, we cannot estimate the accretion rate.
Based on the simulation of \citet{2015MNRAS.453.3213S}, it can be showed that SS 433 has a pair of radiatively driven jets,
which is baryon-loaded, kinetically dominated, and collimated.
For a $10\ M_{\sun}$ BH accreting at $\dot{m}=100$ in SS 433, the jet power is roughly $0.01 \dot{M}c^{2} \approx 2 \times 10^{40} \ \rm{erg \ s^{-1}}$ \citep{2015MNRAS.453.3213S}. 
This is sufficient to explain the kinetic luminosity of the jets observed in SS 433. 
Thus, if the power of the jet in our model is roughly the same as that in the SS 433, we reckon the range of specific accretion rate $\dot{m} = 10-1000$ for the 
binary system of FRB 180916. For an NS-WD binary, as the mass of NS is relatively small, a large accretion rate, $\dot{m} \sim 1000$, is needed to generate the observed radio luminosity. 
For a BH-WD binary, an appropriate specific accretion rate, $\dot{m} \sim 10$, can meet the luminosity criteria because the BH mass is relatively larger than that of an NS.

\begin{figure}[htbp]
\centering
\includegraphics[height=10cm,width=10cm]{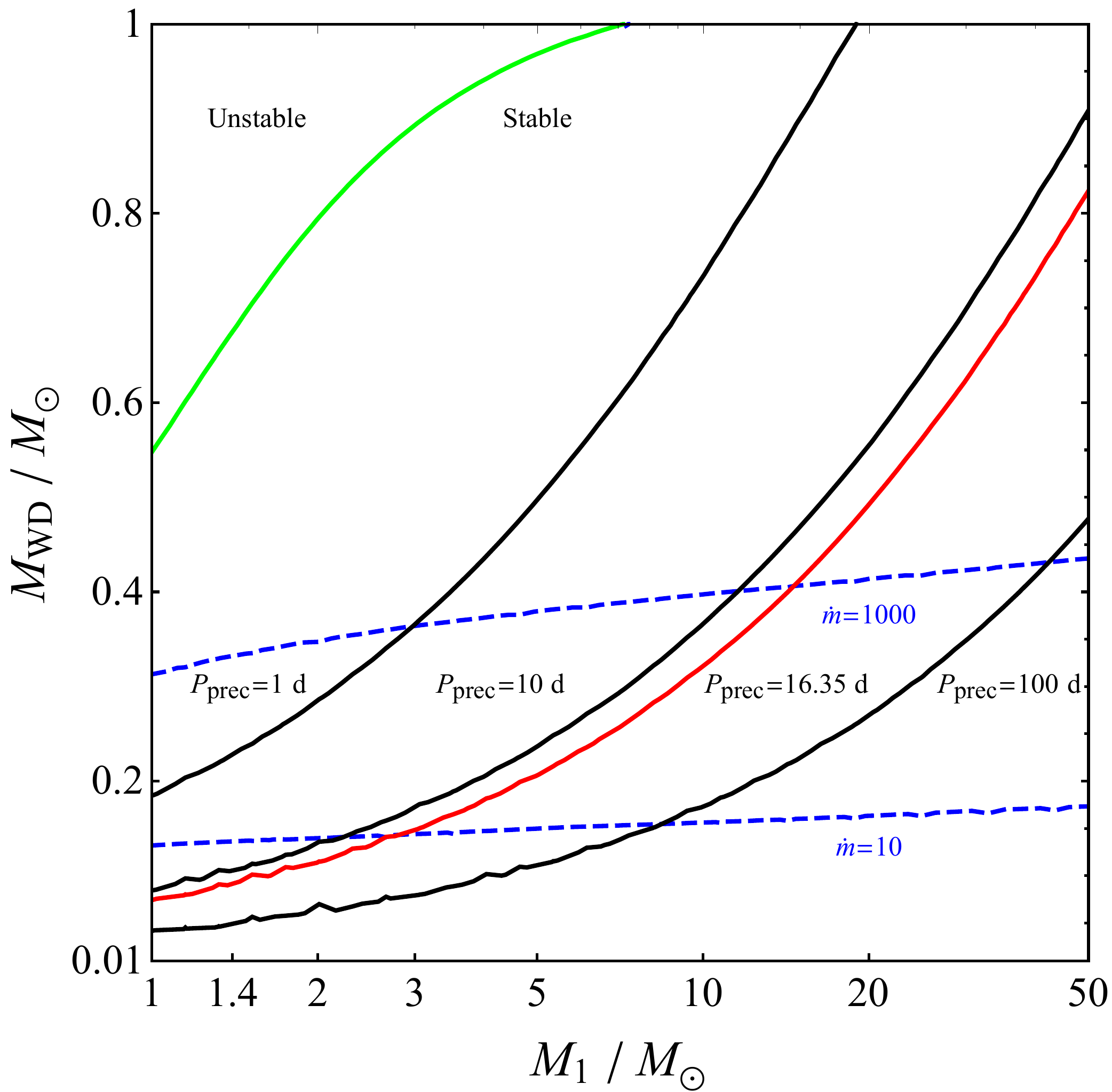}
\caption{The relationship between the precession period $P_{\rm{prec}}$ and the
mass of two stars under the condition of high accretion rates. 
The green line corresponds to the critical mass relationship
between stable and unstable mass transfer.
The blue dashed lines correspond to the specific accretion
rate $\dot{m}=10$ and $\dot{m}=1000$. The black lines correspond to different
precession periods: 1 day, 10 days, and 100 days. The red line
corresponds to the precession period of FRB 180916 (16.35 days). The orbital
inclination with respect to the plane of disk is assumed to be $\theta=20{\degr}$.
\label{fig:period} }
\end{figure}

Combining Equations (\ref{con: Porb/Pprec 2}) and (\ref{con: dot{m}}), 
we can obtain the relationship between the precession period
$ P_{\rm{prec}}$ and the masses of two stars in the range of specific accretion 
rate $\dot{m}=10-1000$ (Figure \ref{fig:period}). 
There is a critical relationship for the mass of two stars 
(the green line in Figure \ref{fig:period}) which corresponds to the boundary between stable and unstable accretion of the binary systems \citep[see Equation (9) in][]{Dong et al.(2018)}. 
Once the masses of the primary stars and WDs beyond the critical mass, 
the unstable and extremely violent accretion will occur \citep{Dong et al.(2018)}. 
Then, the mass transfer process might be accompanied by strong outflows.
The red line in Figure \ref{fig:period} represents the precession period of
FRB 180916. Under the range of $\dot{m}=10-1000$,
the mass of the primary star can be from $M_{1}\sim2.8 \ M_{\sun}$ (a massive NS) to $M_{1} \sim 15 \ M_{\sun}$ (a stellar-mass BH) to reproduce the 16.35-day period in FRB 180916.

In summary, the tidal force driven-jet precession model can explain the periodic activity of FRB 180916 in BH/NS-WD binary system with a high accretion rate. 

\section{Properties of System} \label{sec:system}

\subsection{NS-WD or BH-WD binary } \label{subsec:binary}

If the tidal force-driven jet precession model is correct, we might be able to infer the masses and types of the central stars of other repetitive and periodic FRBs.
Combining Equations (\ref{con: Porb/Pprec 2}) and (\ref{con: dot{m}}),   
if $P_{\rm{prec}} \la 1 \ \rm{day}$, the central objects are massive NSs; if $P_{\rm{prec}} \ga 20 \ \rm{day}$, FRBs can be observed in stellar-mass BH-WD binaries.

\subsection{Duration time of FRB 180916 Generation} \label{subsec:duration}

\begin{figure}[htbp]
\centering
\includegraphics[height=8cm,width=10cm]{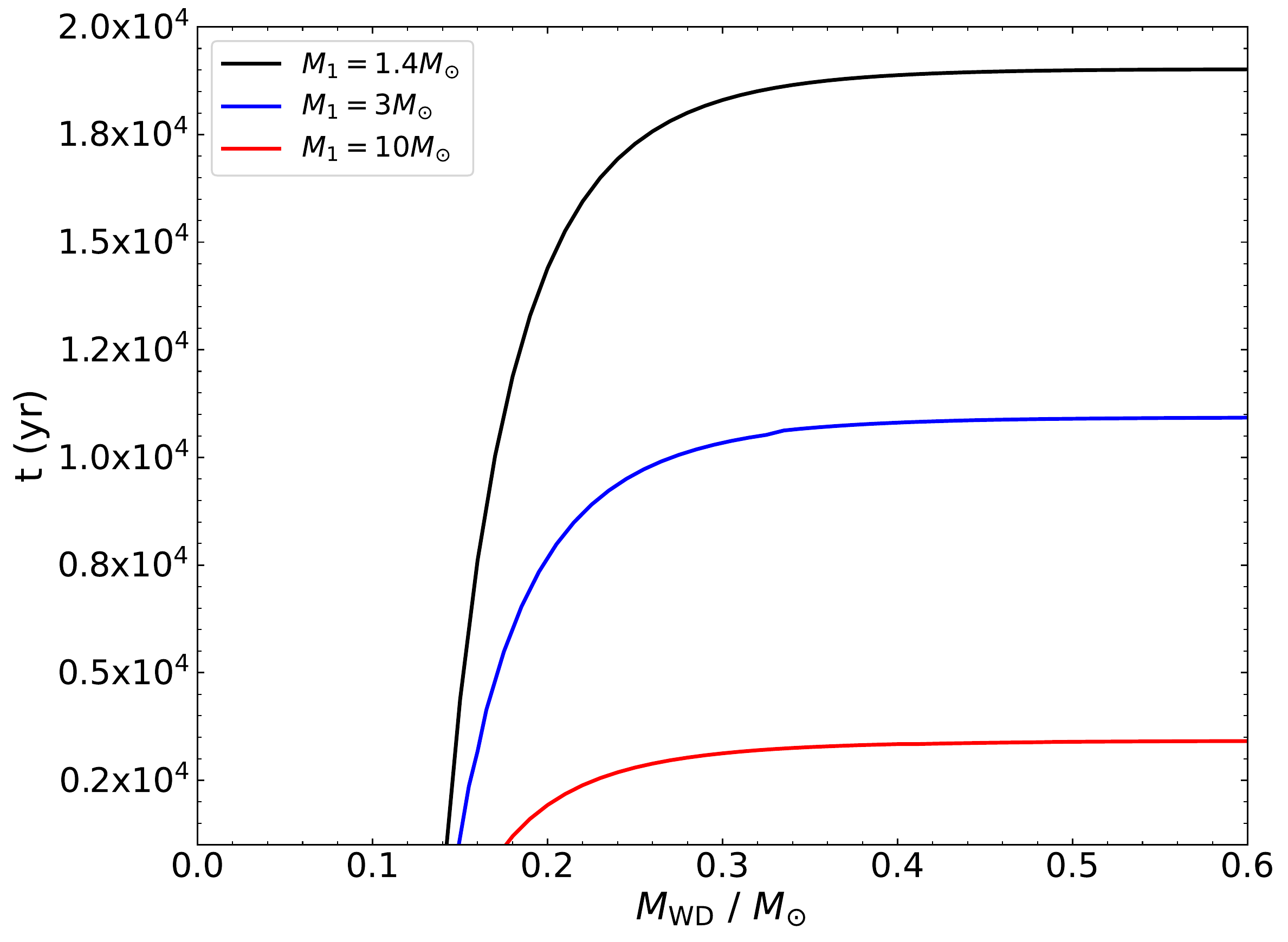}
\caption{The duration timescale of periodic FRBs in binary systems with
high accretion rates. Different lines correspond to different 
mass of the primary star: $1.4 \ M_{\sun}$ (black), $3 \ M_{\sun}$ 
(blue), and $10 \ M_{\sun}$ (red). The duration decreases (increases) with increasing the mass of the primary star $M_{1}$ (the mass of WD $M_{\rm{WD}}$). The intersections of the lines and x-axis are the masses of $M_{\rm{WD}}$ when specific accretion rates $\dot{m}$ are equal to 10, i.e., $\sim 0.14 \ M_{\sun}$ for $M_{1}=1.4 \  M_{\sun}$, $\sim 0.15 \ M_{\sun}$ for $M_{1}=3 \  M_{\sun}$, and $\sim 0.18 \ M_{\sun}$ for $M_{1}=10 \  M_{\sun}$.
\label{fig:time} }
\end{figure}

We can estimate the duration  (denoted as $T$) of a periodic FRB in a binary system with a high accretion rate.
To estimate the duration $T$ of a periodic FRB, we fix $M_{1}=1.4 \  M_{\sun}$ for the NS-WD binary. 
For the BH-WD binary, we fix $M_{1}=3 \  M_{\sun}$ and $10 \ M_{\sun}$. 
In all binaries, the WDs are assumed to have an initial mass of $0.6 \ M_{\sun}$ \citep{Kepler et al.(2007)}.
In our model, the duration $T$ of a periodic FRB is equal to the evolution time of the WD which evolves from the initial mass ($M_{\rm{WD}}=0.6\ M_{\sun}$) to the final mass (e.g., $\dot{m}=10, M_{\rm{WD}}\sim 0.14\ M_{\sun}$; see Equation (\ref{con: dot{m}}) when $M_{1}=1.4\ M_{\sun}$). 
The mass transfer rate from the WD to BH/NS, $\dot{M}=dM_{\rm{WD}}/dT$, can be calculated from Equation (\ref{con: accretion}), where the unknown parameter $a$ depends on $M_{\rm{WD}}$ and $M_{1}$ (see Equations (\ref{con:Rwd}) and (\ref{con:RL2/a})). As $M_{1}$ is usually much larger than $M_{\rm{WD}}$, we can reasonably assume that $M_{1}$ remains constant during the mass transfer process. The time interval between $M_{\rm{WD}}=0.6 \ M_{\sun}$ and $M_{\rm{WD}}=0.14 \ M_{\sun}$ is estimated as follows. 

\begin{equation}
T=\int_{M_{\rm{WD}}=0.14\ M_{\sun}}^{M_{\rm{WD}}=0.6 \ M_{\sun}} \frac{dM_{\rm{WD}}}{\dot{M}}\ .
\end{equation}
The results are shown in Figure \ref{fig:time}.
It shows that the duration $T$ of the periodic FRBs increases (decreases) with increasing $M_{\rm{WD}}$ (the mass of the BH/NS $M_{1}$) for all binaries.
As mentioned above, FRBs can be produced for NS-WD or BH-WD binaries with $\dot{m}>10$. As the specific accretion rate $\dot{m}$ drops below 10, the system might evolve into UCXBs. 
Hence, the duration time of FRB 180916 generation
ranges from thousands of years to tens of thousands of years. 
This duration time is only a small portion of the evolution
time of binaries. 

\subsection{Event Rate} \label{subsec:event}

The event rate of FRBs may be affected by a number of factors, such as redshift, luminosity, and the types of FRBs (repeating or one-off). For example, the volumetric rate of FRBs is roughly $2 \times 10^{3} \ \rm{Gpc}^{-3} \
\rm{yr}^{-1}$ at redshift $z \sim 1$ \citep{Petroff et al.(2019)}, whereas \citet{Cao et al.(2018)} showed that the local event rate is about $(3-6) \times 10^{4} \ \rm{Gpc}^{-3} \ \rm{yr}^{-1}$.
With the luminosity function of 46 known FRBs, the event rate of FRBs was found to be $\sim 3.5\times 10^{4} \ \rm{Gpc^{-3} \ yr^{-1}}$ above $10^{42}\ \rm{erg \ s^{-1}}$, $\sim 5.0\times 10^{3} \ \rm{Gpc^{-3} \ yr^{-1}}$ above $10^{43}\ \rm{erg \ s^{-1}}$, and $\sim 3.7\times 10^{2} \ \rm{Gpc^{-3} \ yr^{-1}}$ above $10^{44}\ \rm{erg \ s^{-1}}$ \citep{Luo et al.(2020)}. 
In addition, \citet{2019ApJ...882..108W} estimated the volumetric rate of repeating FRBs
about $500 \ \rm{Gpc}^{-3} \ \rm{yr}^{-1}$ over $0<z<0.5$. Hence, it seems that most FRBs are non-repeating FRBs.
 
According to our model, the event rate of the periodic FRBs $R_{\rm{FRB}}$ equals to the volumetric rate of BH/NS-WD binary systems with high specific accretion rates $\dot{m} \geq 10$. For the BH-WD binary, according to \citet{Fryer et al.(1999)},
the formation rate of BH-WD binaries is estimated to be $ \sim 0.15 \ \rm{Myr}^{-1} \ 
\rm{galaxy}^{-1}$. Then, we use the relationship between the number density of
galaxy and the age of the Universe \citep{Conselice et al.(2016)} to estimate
the number of galaxies in $0<z<1$. The total number of galaxies within $0<z<1$ can be expressed as \citep{Conselice et al.(2016)}

\begin{equation}
N_{\rm{tot}}= \int_{t_{z=0}}^{t_{z=1}} \int_{0}^{4 \pi} D_{H} \frac{\left(1+z \right)^{2} D_{A}^2}{E(z)} \phi_{T}(t) \ d \Psi d t,
\label{con: number}
\end{equation}
where the volume of the whole sky is interated through $4 \pi$ in steradians, $D_{A}$ is the angular size distance, $D_{H}=c/H_{0} (H_{0}\approx70\ \rm{km} \ {s}^{-1}\ \rm{Mpc}^{-1})$, $E(z)= \left(\Omega_{M} \left(1+z \right)^{3}+\Omega_{\lambda}\right)^{1/2}$($\Omega_{M}\approx0.3,\Omega_{\lambda}\approx0.7$), and $t$ is the age of the universe in units of $\rm{Gyr}$ at redshift $z$. 
The number density of galaxies $\phi_{T}(t)$ can be fitted as \citep{Conselice et al.(2016)}

\begin{equation}
\rm{log}\ \phi_{T}(t)=\left(-1.08 \pm 0.20\right) \times \rm{log} \ t - 0.26 \pm 0.06.
\end{equation}
By multiplying Equation(\ref{con: number})
and the formation rate of BH-WD binaries, we can get the volumetric rate of
BH-WD binaries $\sim 340 \ \rm{Gpc}^{-3} \ \rm{yr}^{-1}$. 
For the NS-WD binary, \citet{Zhao et al.(2021)} calculate the number
of the NS-WD binaries evolved from a unit mass stellar population, $f_{\rm{NSWD}} \sim 2.1 \times 10^{-5} \ M_{\sun}^{-1}$. Then, we need the cosmic star formation rate (CSFR)
to estimate the  volumetric rate of NS-WD binaries.
The CSFR is taken the form \citep{2014ARA&A..52..415M}
\begin{equation}
\psi (z)= 0.015 \ \frac{\left(1+z \right)^{2.7}}{1+\left[\left(1+z \right)/2.9 \right]^{5.6}} \ M_{\sun}\ \rm{yr}^{-1} \ \rm{Mpc}^{-3} \label{con: CSFR}.
\end{equation}
Combining $f_{\rm{NSWD}}$ and Equation (\ref{con: CSFR}), we can evaluate the volumetric rate of NS-WD binaries
in redshift $0<z<1$, i.e., $\sim4500  \ \rm{Gpc}^{-3} \ \rm{yr}^{-1}$.

According to our model, not all BH/NS-WD binaries produce the periodic FRBs. 
Only in BH/NS-WD binaries with high accretion rates ($\dot{m}\geq 10$), can we observe periodic FRBs. Not all BH/NS-WD binaries have mass transfer processes. That is, only a fraction (denoted as $f_{\rm{acc}}$) of BH/NS-WD binaries can power periodic FRBs. Only 14 UCXBs have been confirmed in the Galaxy \citep{van Haaften et al.(2012)}. The donor stars in these binary systems must be WDs or helium-burning stars that fill their Roche lobes since the orbital periods of these binaries are less than one hour. 
Therefore, it seems that $f_{\rm{acc}}$ is very small.
The total number of periodic FRBs should also be proportional to the duration time $T$ and anti-correlate with the jet beaming factor $f_{\rm{b}}$.
Thus, the event rate of the repetitive and periodic FRBs is evaluated to be $\sim 0.05 \ f_{\rm{acc}} \ (f_{\rm{b}}/100)^{-1}\ (T/ 1 \ \rm{kyr}) \ \rm{Gpc}^{-3} \ \rm{yr}^{-1}$.

\section{Conclusions and discussion} \label{sec:con}

In this paper, we have  investigated two precession models to explain the 16.35-day
activity period of FRB 180916: the accretion disk-driven jet precession model
and the tidal force-driven jet precession model. Our main results can be summarized as follows.

\begin{enumerate}
 \item The first model involves the jet precession of a binary system due to the  massive outer portions of the accretion disk. In order to explain the 16.35-day precession period of FRB 180916, an extremely low viscous parameter ($\alpha \sim 10^{-12}-10^{-8}$ depending upon the BH masses) is necessary in either the stellar-mass or intermediate-mass BH accretion system; the required $\alpha$ is implausible (see Figure \ref{fig1}; Section \ref{subsec: jet}).
 \item The 16.35-day period can be well explained by the tidal force-driven jet precession model in either a BH-WD or NS-WD binary with a high accretion rate (see the red line in Figure \ref{fig:period}; Section \ref{subsec: tidal}). 
 \item If the tidal force-driven jet precession model is correct, we might infer the types of the binaries for other repetitive and periodic FRBs from the period. If $P_{\rm{prec}} \la 1 \ \rm{day}$, the binary systems are NS-WD binaries. If $P_{\rm{prec}} \ga 20 \ \rm{day}$, FRBs can be observed in BH-WD binaries (see the black lines in Figure \ref{fig:period}; Section \ref{subsec:binary}). 
 \item As assumed in the tidal force-driven jet precession model, the repeating FRBs can be produced by the precession jet in BH/NS-WD binaries with a stable and super-Eddington accretion. If the accretion state is violent and extremely super-Eddington ($\dot{m}>10^{4}$), a BH/NS-WD binary might manifest itself as a long GRB \citep{Dong et al.(2018)}. As the accretion rate declines, repeating FRBs might further evolve into UCXBs.
(see Section \ref{subsec:duration}).
  
\end{enumerate}

FRB 121102 was reported to have a 157-day period \citep{Rajwade et al.(2020)}.
If the periodic activity of FRB 121102 is driven by  
the tidal force-driven jet precession, the binary system is likely a BH-WD binary. 

In the activity cycle, FRB 180916 has a 5-day (a duty cycle of $\sim$ 30$\%$) 
burst window \citep{CHIME/FRB Collaboration et al.(2020a)}
and FRB 121102 has a long duty cycle of $\sim$ 56$\%$ \citep{Rajwade et al.(2020),Cruces et al.(2021)}. 
In our model, the duty cycle of the activity
period may be related to the viewing angle of the observers $\theta_{\rm{obs}}$ and
the half opening angle of the precession cone $\theta_{\rm{c}}$. When
$\theta_{\rm{obs}}<\theta_{\rm{c}}$, we can observe FRBs about in the half
of the precession period which indicates a long duty cycle, e.g., FRB 121102. When $\theta_{\rm{obs}}>\theta_{\rm{c}}$, we can only 
observe FRBs in a small part of the precession cycle which indicates a short
duty cycle, e.g., FRB 180916.

A BH/NS-WD binary with a high accretion rate should have a strong X-ray radiation. However, since FRBs are far from us, e.g., the distance of FRB 180916 is 149 Mpc \citep{Marcote et al.(2020)}, we may not receive any X-ray radiations. 

The BH/NS-WD binaries can emit gravitational wave radiation. The period of such systems directly coincides with the LISA observation band $(10^{-4}-10^{-1}\ \rm{Hz})$. In the future,  the connection between FRBs and gravitational waves may be confirmed.

\acknowledgments

We thank Xiang-Dong Li and Shan-Shan Weng for helpful discussion, and thank the referee for constructive suggestions that improved  the manuscript.
This work was supported by the National Natural Science Foundation of China
under grants 11925301, 12033006, 11822304, and 11973002.

\end{document}